\documentclass[twocolumn]{aastex701}
\usepackage{amsmath} 
\usepackage{graphicx}
\usepackage{nicefrac}
\usepackage{orcidlink}

\begin{document}

\title{Evolution of Main Sequence Stars Transferring Mass to a Supermassive Black Hole}

\author[orcid=0000-0003-3932-2544]{Andrey Sandomirsky}
\affiliation{Racah Institute of Physics, The Hebrew University of Jerusalem, 9190401, Israel}
\email[show]{andrey.sandomirsky@mail.huji.ac.il} 

\author[orcid=0000-0002-1084-3656]{Re'em Sari}
\affiliation{Racah Institute of Physics, The Hebrew University of Jerusalem, 9190401, Israel}
\email{sari@phys.huji.ac.il}

\author[orcid=0000-0002-6105-6492]{Aleksandra Olejak}
\affiliation{Max Planck Institute for Astrophysics, Karl-Schwarzschild-Straße 1, 85748 Garching b. München, Germany}
\email{aleksandra.olejak@wp.pl}

\author[orcid=0000-0001-9336-2825]{Selma E. de Mink}
\affiliation{Max Planck Institute for Astrophysics, Karl-Schwarzschild-Straße 1, 85748 Garching b. München, Germany}
\email{sedemink@mpa-garching.mpg.de}

\begin{abstract}
We consider main-sequence stars orbiting close to a supermassive black hole (SMBH), which are potential gravitational wave (GW) sources for LISA if their orbital periods are of the order of an hour. At such a short orbital period, mass transfer from the star to the SMBH occurs. The evolution of the semi-major axis and GW frequency depends on the evolution of the stellar mass and radius. We use MESA to study stars that transfer mass to SMBHs similar to Sagittarius A* starting on the zero-age main sequence (MS). We identify 4 evolutionary phases. (I) Stars initially with mass $>2 M_\odot$ remain on the MS as their mass and radius decrease. (II) Below $2~M_\odot$, the separation is sufficiently small so that the GW timescale is too short for the stars to maintain thermal equilibrium. They evolve adiabatically off the MS and shrink rapidly as they lose their high-entropy envelope. (III) Below $0.5-1~M_\odot$, depending on the initial mass, the uniform low-entropy core is exposed and stars expand adiabatically. (IV) Below $0.15-0.4~M_\odot$, the thermal and GW timescales become comparable, and stars cool and shrink while maintaining this balance. Overall, mass transfer causes the orbit to expand, slowing down the orbital evolution and leading to GW emission at lower frequencies making the GW signal harder to detect. If located around Sagittarius A*, mass transferring stars spend most of the time relatively close to the LISA sensitivity curve, with maximal SNR reaching around 600 during the transition between stages II and III.
\end{abstract}


\section{Introduction}

Many galactic centers host a massive black hole surrounded by a nuclear stellar cluster \citep{Volonteri2010}. Dynamical processes in such a dense stellar environment lead to a variety of phenomena from stellar collisions \citep{Balberg2023, Balberg2024, Rose2025} to emission of gravitational waves (GWs) \citep{Freitag2003, AmaroSeoane2012, AmaroSeoane2012eLISA, Gair2013, AmaroSeoane2007, AmaroSeoane2017, Rom2024} and tidal disruption events (TDEs) \citep{Hills1975, Rees1988, Evans1989, Piran2015, Krolik2016, Dai2018, Jonker2021, Rom2025}. The latter is a violent event, where a star on an extremely eccentric orbit arrives close enough to a supermassive black hole (SMBH), closer than the tidal radius, and 
dissolves into streams of gas. This phenomena is believed to have been seen in over a hundred of distant galaxies \citep{Gezari2006, Gezari2008, Gezari2021, vanVelzen2021}. The changing tidal potential of a SMBH for stars on eccentric orbits and their proximity to it alter their internal structure, potentially leading to premature envelope loss via episodic mass transfer to the SMBH \citep{Guillochon2013, MacLeod2013, Ryu2020, Bush2025}.
There is, however, a theoretically predicted less violent class of TDEs that involves a star on an almost circular orbit with a semimajor axis just equal to its tidal radius, that is slowly shedding mass. This process is closely related to mass transfer in binaries in general \citep{Paczynski1971}.
This phenomena is the focus of this paper.

When a star overfills its Roche lobe,  matter flows through the inner Lagrange point, a process known as Roche-lobe overflow (RLOF) \citep{Ritter1988, Kolb1990}. Mass transfer may influence stellar structure and evolution and is governed by several effects. One is that the donor star might contract in response to mass loss, which helps it to remain within its Roche lobe and to stabilize the mass transfer \citep{Rappaport1982}. Another one is the transfer of angular momentum during mass exchange. When mass flows from a heavier donor to a lighter companion, the binary orbit tends to shrink \citep{Tauris2023}. If the donor is lighter, the orbit usually widens, and we are interested in this case. The orbital expansion can enlarge the Roche lobe or slow its contraction, stabilizing the mass transfer process.

Whether the mass transfer is stable depends on how the donor star's radius and Roche lobe respond to mass loss \citep{Dai2013b, Pavlovskii2015, Pavlovskii2016}. If the star expands more quickly than its Roche lobe or does not shrink fast enough, the process becomes unstable and may lead to the destruction of the star. Conversely, if the star contracts faster or expands more slowly than its Roche lobe, the mass transfer tends to remain stable and can continue for an extended period \citep{Soberman1997}.

In this paper, we are interested in a main-sequence (MS) star orbiting close to a supermassive black hole (SMBH). Such binaries are gaining recent interest due to their possible relation with QPEs - Quasi Periodic X-ray Eruptions observed from galactic centers \citep{Giustini2020, Arcodia2021, Chakraborty2021, Krolik2022, Miniutti2023, Linial2023}.
Stars occupying circular orbits slowly spiral in due to the emission of gravitational waves, and eventually start to transfer mass to the SMBH \citep{Miller2005, AmaroSeoane2012, Linial2017, Metzger2017, Metzger2022}. Besides QPEs, a range of other observational phenomena may provide indirect evidence for the presence of stars orbiting SMBHs on mildly eccentric or nearly circular trajectories. In particular, a mechanism involving such systems has been proposed to explain jetted TDEs \citep{Berger2012, LinialQuataert2024,Linial2026}, a rare subclass of TDEs characterized by the launch of powerful relativistic jets evidenced by bright non-thermal X-ray and radio emission. The nuclear transient Swift J0230, which exhibited recurrent X-ray flares with a period of 22 days, has been suggested to originate from a giant star transferring mass onto an SMBH \citep{Guolo2024}. Mass transfer from stars onto SMBHs has long also been considered a potential contributor to the fueling of active galactic nuclei \citep{Hameury1994}.
While TDEs occur on the timescales of several hours, mass transfer from stars revolving around BHs on circular orbits lasts for millions of years \citep{Dai2013b, Guillochon2013, Stone2019,LinialQuataert2024, Olejak2025}. The interaction is governed by the balance of forces of stellar self-gravity and the tidal field imposed by the SMBH. 

The binary systems in which a stellar-mass object gradually inspirals into a SMBH through the emission of GWs are so called extreme mass ratio inspirals (EMRIs) \citep{AmaroSeoane2007}.
The stellar mass object is usually taken to be a stellar-mass black hole \citep{Alexander2005, Hopman2006},
and such systems are promising sources for the future mission Laser Interferometer Space Antenna (LISA)  \citep{AmaroSeoane2007, AmaroSeoane2017, Rom2024}.
However, main sequence stars are more numerous than stellar mass BHs, and also approach the SMBH by emission of gravitational waves. This eventually leads to mass transfer. Mass transfer affects the orbital evolution and therefore the gravitational wave signal. Thus, modeling of the stellar mass loss and orbital evolution is essential to predict the gravitational signals that can be detected in LISA band from main sequence stars.

Our goal is to investigate  how conservative mass transfer modifies the mass-radius relation of a MS star and its inspiral into a SMBH by combining stellar evolution models with orbital change calculations. This allows us to estimate the gravitational waves radiation which can be detected by future instruments like LISA.

The paper is structured as follows. In Section $\mathsection$\ref{sec: Methods. Mass transfer} we characterize mechanism of mass transfer from a star to a SMBH and derive stability criterion for it. In Section $\mathsection$\ref{sec: MESA code} we describe setup for our simulations in MESA code. In Section $\mathsection$\ref{sec: Methods. Timescales and orbital evolution} we introduce relevant timescales and present binary separation and GW evolution from MESA. In Section $\mathsection$\ref{sec: Results. Radius-Mass relation} we present results for mass-radius relation as well as entropy profiles and analyze them. In Section $\mathsection$\ref{sec: Results. GW radiation} we discuss resulting GW signals and LISA capabilities of detecting them. Finally, in Section $\mathsection$\ref{sec: Summary and Discussion} we summarize our findings and discuss further prospectives on the topic.

\section{Mass Transfer}
\label{sec: Methods. Mass transfer}
We consider a binary system of a primary (a SMBH of a mass $M$) and a secondary (a star of a mass $m$) on a circular orbit with a semi-major axis (a binary separation) $a$. In such a system, the teardrop-shaped area around each of binary components where orbiting material is gravitationally bound to that specific component is called a Roche lobe.

The Roche lobe radius $R_\text{RL}$ -- an effective radius of the Roche lobe around the secondary, is defined as the radius of a sphere that has the same volume as the Roche lobe. It depends on the mass ratio $\frac{m}{M}$, and in the case of SMBH, when condition $m \ll M$ holds, is approximated by the low mass ratio limit of the Eggleton formula \citep{Eggleton1983}:

\begin{equation}
    R_\text{RL} = 0.49a{\left(\frac{m}{M}\right)}^{\nicefrac{1}{3}}.
    \label{eq:Eq1}
\end{equation}
Mass transfer from the star to the SMBH occurs when the stellar radius $R$ exceeds $R_\text{RL}$.
We assume mass conservation -- all the mass that was lost by the star goes directly to the SMBH: $\dot{M} = -\dot{m}$.

The relation between $a$ and binary orbital period $T$ is given by the Kepler's third law:
\begin{equation}
    {\left(\frac{2\pi}{T}\right)}^2 = \frac{G(m + M)}{a^3} \ .
    \label{eq:3rdKepler}
\end{equation}

In this work we consider a system where the total angular momentum is conserved: $L=\text{const}$.
As stellar material is accreted onto the SMBH, it spirals in from larger to smaller orbits, thereby losing its angular momentum. In a closed system, this angular momentum must be transferred back to the star in order to conserve the total angular momentum. The fraction of angular momentum that could end up in the SMBH is small, since its innermost stable circular orbit (ISCO) is much smaller than the star's orbital radius. The exchange is governed by tidal torques between the star and the SMBH’s accretion disk, which return angular momentum from the disk to the star. We assume that angular momentum does not accumulate significantly in the accretion disk. Equivalently, this corresponds to the disk remaining less massive than the star.

Effects of general relativity (GR) except gravitational waves are negligible (Subsection $\mathsection$\ref{subsec: Methods. GW timescales}, Appendix \ref{sec: Change of angular momentum and equation of motion}), since we study binary systems with a semi-major axis significantly larger than the Schwarzschild radius of the SMBH (the initial binary period is 1 day, Section $\mathsection$\ref{sec: MESA code}).

The total angular momentum of the system:

\begin{equation}
    L = \mu\sqrt{G(m + M)a} \ ,
    \label{eq:Ang. momentum of the binary}
\end{equation}
where $\mu = mM/(m + M)$ -- the reduced mass.

Writing $\dot{L}/L$ and using Eq. \ref{eq:Ang. momentum of the binary}, we derive the equation of motion (EoM) in $m \ll M$ limit: 

\begin{equation}
    \frac{\dot{a}}{a} = -2\frac{\dot{m}}{m} \ .
    \label{eq:Eq9}
\end{equation}

Let us denote radius-mass dependence for the star:
$R \propto m^\epsilon$ where $\epsilon$ changes depending on different regimes of stellar evolution.
Mass transfer is stable if for a  binary that evolves under mass transfer only, the mass transfer rate decreases with time. Thus, stability requires that the time derivative of the ratio $R/R_\text{RL}$ be negative (using $m \ll M$ and $\dot{M} = -\dot{m}$):

\begin{equation}
\begin{split}
    \frac{d}{dt}\left(\frac{R}{R_\text{RL}}\right) = \frac{\dot{m}}{m}\left(\epsilon + \frac{5}{3}\right) < 0 \ .
    \label{eq:Eq10}
\end{split}
\end{equation}
Since $\dot m<0$, the stability criterion is
\begin{equation}
    \epsilon + \frac{5}{3} > 0 \ .
    \label{eq:Eq12}
\end{equation}

\section{Our MESA setup}
\label{sec: MESA code}

In this work, we compare our theoretical predictions with numerical simulations. For this purpose, we employ the stellar evolution code MESA (Modules for Experiments in Stellar Astrophysics) \citep{Paxton2010, Paxton2013, Paxton2015, Paxton2019}.

We use the "star\_plus\_point\_mass" module. This module models a binary system where one component is a MS star and the other is treated as a not evolved point mass, representing, in our case, a SMBH. The gravitational influence of the point mass is included in the orbital evolution, but its internal structure is not modeled -- it acts purely as a sink for mass.

Within the ``star\_plus\_point\_mass" module, mass transfer occurs if the donor star overfills its Roche lobe. 
The code computes the Roche lobe radius using the approximation from 
\citet{Eggleton1983}, and adjusts the mass transfer rate based on the star's radius relative to the Roche lobe. Orbital evolution due to gravitational radiation and mass loss are also self-consistently modeled. This simulation is based on the angular momentum conservation model (Section $\mathsection$\ref{sec: Methods. Mass transfer}) where the binary's angular momentum changes only due to GW radiation.

We run stellar models with initial masses in the range $0.5-10$ solar masses with respectively different radii, initially on the MS. These stars orbit a SMBH with $M=10^6~M_\odot$, on an initial orbit with a period of one day, equivalent to a semi-major axis of $420$ solar radii. This separation ensures that no mass transfer occurs initially, although the star begins to overfill its Roche lobe soon after the start of the simulation (after $\sim 1.9 \cdot 10^5$~years), before any significant stellar evolution occurs. The mass transfer in such a setup is always stable according to Eq.~\ref{eq:Eq12}.
The GW emission controls how quickly the stars approach the SMBH and when mass transfer sets in. We tracked stellar structure, mass loss rate, and orbital decay throughout the simulations.

Our use of this module allows us to automatically track stellar evolution while taking into account the relation between the mass loss timescale and the thermal timescale of the star. This is different from the approach of \citet{Dai2013b}
 (see also the analytic treatment of \citet{Paczynski1965}) who calculated adiabatic evolution. 

\section{Timescales and orbital evolution}
\label{sec: Methods. Timescales and orbital evolution}

\subsection{Kelvin-Helmholtz timescale}
\label{subsec: Methods. KH timescale}

The Kelvin-Helmholtz (KH) or thermal timescale is the time it takes for a star to radiate all its thermal energy at its current luminosity. It is estimated as gravitational potential energy divided by the stellar luminosity $L$:

\begin{equation}
    t_\text{KH} \sim \frac{Gm^2/R}{L} \ .
    \label{eq: t_KH}
\end{equation}

We assume a star is a spherically symmetric body that radiates energy as a black body, then:

\begin{equation}
    t_\text{KH} \sim \frac{Gm^2/R}{(\frac{4\sigma}{c}T^4\cdot\frac{4}{3}\pi R^3)/(\frac{R^2}{lc})} \ ,
    \label{eq:t_KH estimate from theory}
\end{equation}
where $\sigma$ is the Stefan-Boltzmann constant, $T$ -- the stellar temperature, $l=1/\kappa\rho$ -- the mean free path, $\kappa$ -- the effective opacity of the relevant layers, $\rho$ -- the stellar mass density.

We assume that during the evolution the star stays in hydrostatic equilibrium (HSE) -- balance between forces of gas pressure and gravity. Then we can roughly estimate the temperature in the center of the star:

\begin{equation}
    \frac{Gmm_\text{p}}{R} \sim k_\text{B} T \rightarrow T \propto \frac{m}{R} \ .
    \label{eq:Eq26}
\end{equation}
where $m_\text{p}$ is the proton mass, $k_\text{B}$ - the Boltzmann constant.
Subsequently, the Kelvin-Helmholtz time scales as follows:

\begin{equation}
    t_\text{KH} \propto \frac{\kappa}{mR} \ .
    \label{eq:t_KH scaling}
\end{equation}
For main sequence stars, with $m \gtrsim M_\odot$, the opacity is dominated by Thomson scattering and is constant, hence, $t_\text{KH}$ is short for heavy stars and long for low-mass stars.

\subsection{Gravitational waves timescale}
\label{subsec: Methods. GW timescales}

Gravitational waves (GW) timescale is the time that would take gravitational waves to bring binary system all the way to merger (if there is no mass transfer). In other words, it is the orbital energy of a secondary around a primary divided by GW luminosity.

The orbital evolution equation of the binary radiating gravitational waves is \citep{Peters1964}:
\begin{equation}
    \frac{a}{\dot{a}}\Big|_{\text{GW}} = \frac{1}{2}\frac{L_\text{GW}}{\dot{L}_\text{GW}} = - \frac{5}{64}\frac{c^5}{G^3}\frac{a^4}{mM(m + M)} \ .
    \label{eq:Eq28}
\end{equation}

\noindent Solving this differential equation and expressing $t(a)$ from it gives the timescale of gravitational waves -- the inspiral time (or the binary merger time):

\begin{equation}
    t_\text{GW} = \frac{5}{256}\frac{c^5}{G^3}\frac{a^4}{mM(m + M)} \ .
    \label{eq:Eq30}
\end{equation}

\noindent If the BH is supermassive ($M \gg m$), then $m + M \approx M$. We can derive the semi-major axis for a Roche lobe overfilling star from Eq.~\ref{eq:Eq1}: $a = 2.04R{(M/m)}^{\nicefrac{1}{3}}$. Then from Eq.~\ref{eq:Eq30} we obtain:

\begin{equation}
    t_\text{GW} \approx \frac{87}{256}\frac{c^5}{G^3} \frac{R^4}{m^{7/3}M^{2/3}} \ .
    \label{eq:Eq31}
\end{equation}
As we show later (see \S\ref{sec: Results. Radius-Mass relation}), once mass transfer starts, this GW timescale is closely related to, and about an order of magnitude smaller than, the mass transfer timescale $t_{\dot m}$. Note that the mass transfer rate $\dot{m} < 0$, but in the  MESA output data its absolute value $|\dot{m}|$ is provided, so further we treat $t_{\dot m}$ as a positive quantity:
\begin{equation}
    t_{\dot m} \equiv {m \over |\dot m|} \ .
\end{equation}

\subsection{Binary separation and GW frequency evolution}
\label{subsec: Methods. a and f_GW}

\vspace{0.5mm} 
\begin{figure} [htbp]
\centering
\includegraphics[width=1.\linewidth]{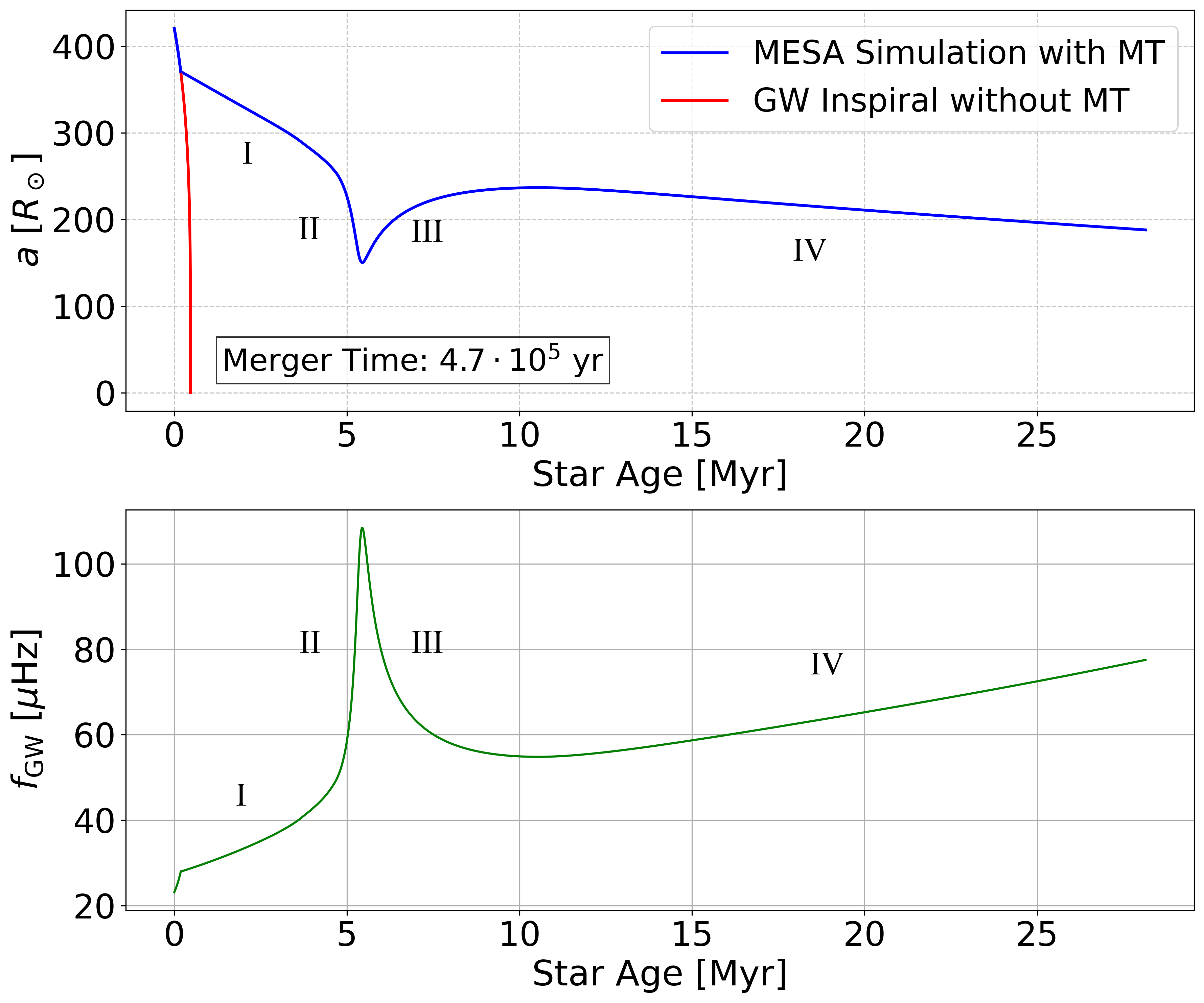}
\caption{Evolution of initially $10~M_\odot$ star orbiting the $10^6M_\odot$ SMBH with $T_0 = 1$ day and $a_0 \approx 420~R_\odot$. {\it Upper panel:} the semi-major axis as a function of the star age $a(t)$. The red line is the evolution of a point mass around the SMBH according to Eq.~\ref{eq:Eq30}, the corresponding merger time is $4.7 \cdot 10^5$~years. The blue line is the result of MESA. The two are identical until mass transfer (MT) starts. Then, conservative mass transfer causes the orbit of the star to shrink more slowly, and even increase. {\it Lower panel:} the GW frequency as a function of the star age $f_\text{GW}(t)$. At stage I the frequency grows moderately when the star evolves along the MS. At stage II the frequency increases sharply and reaches the maximum at the minimal binary separation ($1.5~M_\odot \lesssim m \lesssim 2~M_\odot$). After the peak the frequency decreases at stage III since star is pushed away from the SMBH by mass transfer and then increases again but more gradually at stage IV.}
\label{fig:Fig5}
\end{figure}

The evolution of the binary separation without mass transfer from Eq.~\ref{eq:Eq30} (the red line), binary separation with mass transfer (the blue line) and the GW frequency (the green line) for initially $10~M_\odot$ star from MESA are presented on Fig.~\ref{fig:Fig5}. In the case of two point masses without mass transfer, the binary separation evolves during the inspiral time. In the case of the star and the SMBH, at the separation $a \approx 371~R_\odot$ (after $\sim 1.9 \cdot 10^5$~years of the binary evolution), the star starts to overfill its Roche lobe and mass transfer begins.
It slows down the evolution and leads to more gradual shrinkage of the binary orbit at stage I.
At stage II, starting from $\sim 2~M_\odot$, we observe very steep drop of the binary separation due to different hierarchy of timescales that we describe in Section $\mathsection$\ref{sec: Results. Radius-Mass relation}. At stage III the orbit increases steep up to $\sim 0.5~M_\odot$. Finally, at stage IV the semi-major axis starts to decrease again but more moderately in comparison with $> 2~M_\odot$ region.

We notice increasing of the $f_\text{GW}$ at stage I and sharp growth at stage II ($0.5~M_\odot \lesssim m \lesssim 2~M_\odot$) (what corresponds to decreasing and sharp drop of $a$ respectively) with the maximum at the minimal binary separation. Then at stage III star moves away and frequency decreases (what corresponds to the increasing of $a$). Finally, at stage IV star again approaches the SMBH but more gradually and the frequency again starts to increase.

\section{Radius Evolution With Mass Loss}
\label{sec: Results. Radius-Mass relation}

\begin{figure} [htbp]
\centering
\includegraphics[width=1\linewidth]{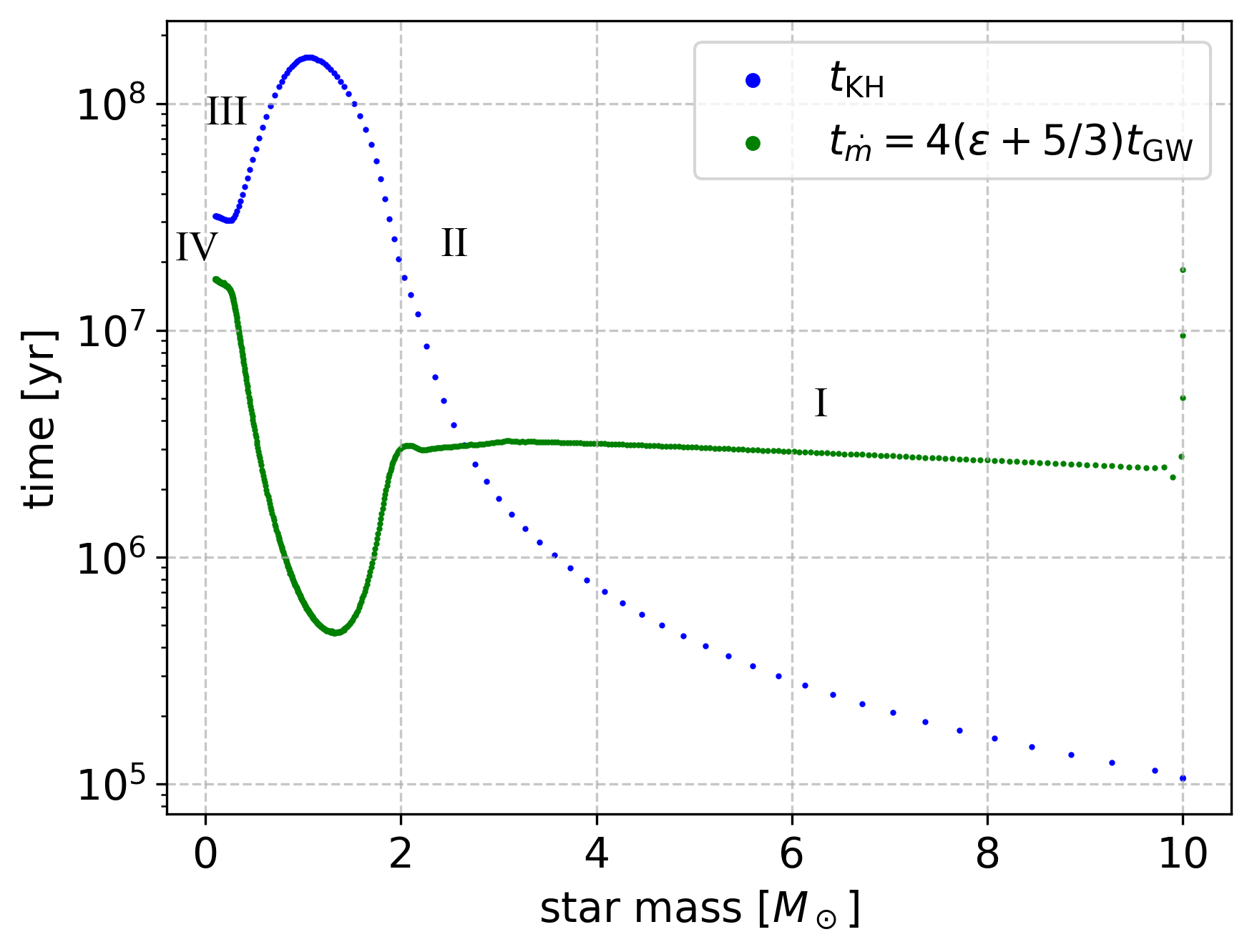}
\caption{$t_{\dot m}$ and KH timescales from MESA for initially $10~M_\odot$ star ($M = 10^6~M_\odot$, $T_0 = 1$ day, evolution before $0.1~M_\odot$).}
\label{fig:GWandKH}
\end{figure}

Fig.~\ref{fig:GWandKH} shows the timescale of the mass transfer from the initially $10~M_\odot$ star to the SMBH ($t_{\dot m}$) and the Kelvin-Helmholtz timescale obtained from the MESA simulation. The characteristic time of GW radiation in MESA is better described by $t_{\dot m}$ because it takes into account radius-mass dependence of the star ($R \propto m^{\epsilon}$). The ratio between $t_{\dot m}$ and time of GW radiation is derived from the EoM (Appendix \ref{sec: Change of angular momentum and equation of motion}):

\begin{equation}
    \frac{t_{\dot m}}{t_\text{GW}} = 4\left(\epsilon + \frac{5}{3}\right).
    \label{eq:Eq72}
\end{equation}
We checked this ratio by plotting  $t_{\dot m}$ from MESA together with $4\left(\epsilon + \frac{5}{3}\right)t_\text{GW}$ using $t_\text{GW}$ from Eq.~\ref{eq:Eq30} and taking $a$, $m$ and $M$ from MESA. They matched each other.
The equality of KH time to GW timescale corresponds to stellar mass of $3.9~M_\odot$ in order of magnitude estimate from  \citet{Linial2017}. Our more precise result from MESA gives $2.6~M_\odot$.

In our exploration, the mass transfer is fast enough, necessitating significant Roche lobe overfilling. Therefore, mass transfer occurs not only through L1, but also through L2 (see the criterion in \citet{Ryu2025}). The mass transfer rates through the two Lagrange points are approximately the same due to extreme mass ratio. We also implemented simulations with non-conservative mass transfer when half of the transferred mass is lost from the vicinity of the star and only half of the mass reaches the SMBH (other initial parameters were the same: $m_0 = 10~M_\odot$, $M = 10^6~M_\odot$, $T_0 = 1$ day). This causes faster evolution of the star in time and the relation between mass transfer timescale and GW timescale  is now $t_{\dot m}/t_\text{GW} = 4\left(\epsilon + \frac{2}{3}\right)$ (derived in the same manner as Eq.~\ref{eq:Eq72}). In this case mass-radius relation approximately stays the same and mass transfer timescale is approximately 2 times shorter. The equality of KH time to GW timescale in this non-conservative MESA run is reached at $\sim 3.4~M_\odot$.

\begin{figure} [htbp]
\raggedright
\includegraphics[width=1.\linewidth]{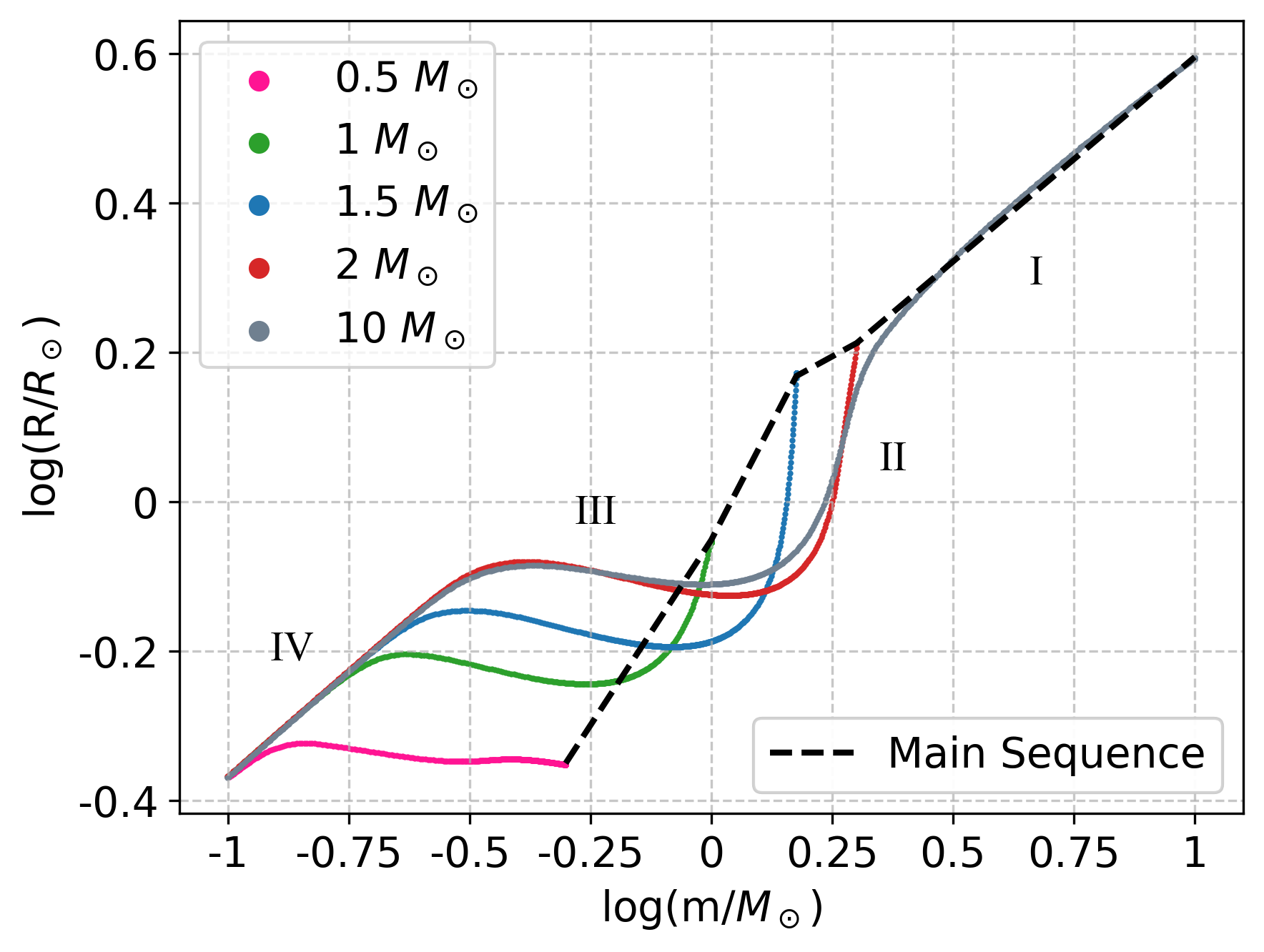}
\caption{Radius-mass relation (in logarithmic scale) of initially $0.5, 1, 1.5, 2, 10~M_\odot$ stars at different stages of their evolution. In these five separate simulations the SMBH mass was $M = 10^6~M_\odot$ and the initial binary period was $T_0 = 1$ day. The end of the simulations was when the mass dropped below $0.1~M_\odot$. Above two solar masses, the mass-radius relation overlap, following that of MS stars (stage I). Once the star reaches $2~M_\odot$, it starts to shrink adiabatically (stage II) and then to expand adiabatically (stage III). Below $2~M_\odot$, all evolutionary tracks starting with more than $2~M_\odot$ overlap, but those with initially less than $2~M_\odot$ differ from each other. Below $\sim 0.5~M_\odot$ stars cool down and gradually shrink at $t_\text{KH} \approx t_\text{GW}$ and the radius-mass evolutionary tracks converge again to a universal behavior (stage IV).}
\label{fig:Fig13}
\end{figure}

We show results for 5 separate simulations for a binary system of a star and a SMBH with different initial masses on Fig.~\ref{fig:Fig13}. We estimated the mass-radius relation exponent $\epsilon$ on this graph. Stars with initial mass above $\sim 2~M_\odot$ evolve along the MS and their mass-radius relation overlap ($R \propto m^{0.6}$, stage I). For instance, for initially $10~M_\odot$ star, above two solar mass $\epsilon$ is in the range $0.5<\epsilon<1.7$ and therefore the ratio $t_{\dot m}/t_\text{GW} \approx 9 - 14$. 

We notice deviation from the MS below $\sim 2~M_\odot$ -- a steep drop of the radius as a function of mass (stage II). This occurs because $t_\text{KH}$ becomes too long compared to the system evolution time  which is given by the gravitational wave timescale. Therefore, is not enough time for the star to adjust itself to thermal equilibrium. 
For initially $10~M_\odot$ star, over the steep decline of $R(m)$ between $\sim 2~M_\odot$ and $\sim 1.5~M_\odot$, on average $R \propto m^{2.3}$ with maximal $\epsilon = 2.7$, hence maximal $t_{\dot m}/t_\text{GW} \approx 17$. 
 
Previous mass transfer investigations \citep{Dai2013, Linial2017} also concluded that for low mass stars the mass-radius curve would deviate from that of the MS, and indeed our simulation confirms this. However, the deviation is in the opposite direction to that found in \citet{Linial2017} and in the same direction as in \citet{Dai2013b}: we find that stars of low mass shrink drastically when losing mass rather than expand (stage II). 
We find that the radius shrinks by more than a factor of two, while the mass drops by $\sim 25~\%$.
For stars with initial mass more than $\sim 2~M_\odot$, this steep radius drop starts at a mass $\sim 2~M_\odot$ and ends at a mass of $\sim 1.5~M_\odot$ (the stage duration $\sim 4 \cdot 10^5$~years).
For stars that start to evolve with $m_0 \lesssim 2~M_\odot$, this sharp radius decrease starts also at $\sim 2~M_\odot$ and ends at minimum at $\sim 1~M_\odot$ (the stage duration $\sim 6 \cdot 10^5$~years).

Farther evolution below $\sim 1.5~M_\odot$ (for stars with initial mass more than $\sim 2~M_\odot$) and below $\sim 1~M_\odot$ (for stars with initial mass less than $\sim 2~M_\odot$) shows an adiabatic expansion of the star in agreement with the prediction  of \citet{Linial2017} (stage III, for initially $10~M_\odot$ $\epsilon$ is in the range $-0.2<\epsilon<0$).
Finally, below $\sim 0.4~M_\odot$ for $m_0 \gtrsim 2~M_\odot$ (below $\sim 0.25~M_\odot$ for $m_0 = 1.5~M_\odot$, below $\sim 0.19~M_\odot$ for $m_0 = 1~M_\odot$, below $\sim 0.13~M_\odot$ for $m_0 = 0.5~M_\odot$) stars start to cool down and shrink again but more moderate on $t_\text{KH} \approx t_{\dot m}$ (stage IV, $R \propto m^{0.6}$), and they already evolved off the MS.

To understand this steep decline, and the difference between our work and the previous analysis which concluded that expansion may occur, we first note the structure of a $2~M_\odot$ star.
MS stars with $m \gtrsim 1.2~M_\odot$ have a convective core encompassed by a radiative envelope which has a higher entropy \citep{Kippenhahn1994}. For two solar mass MS stars, the mass of the convective core is about $1.6~M_\odot$.

In the analysis of \cite{Linial2017} a star is modeled as an object with a uniform entropy. Then in the case of adiabatic evolution:

\begin{equation}
    \rho \sim \frac{m}{R^3}, P \propto {\rho}^{\gamma} \rightarrow P \propto {\left(\frac{m}{R^3}\right)}^{\gamma} \ .
    \label{eq:Eq67}
\end{equation}
HSE then implies:
\begin{equation}
    \frac{Gm^2}{R^4} \sim P \rightarrow R \propto {m}^{\frac{2-\gamma}{4-3\gamma}} \ .
    \label{eq:Eq68}
\end{equation}
In the approximation of a monoatomic ideal gas, for $\gamma = 5/3$:
\begin{equation}
    R\propto {m}^{-\nicefrac{1}{3}} \ ,
    \label{eq: R-m dependence -1/3}
\end{equation}
so as star loses its mass $m$, its radius $R$ grows what determines adiabatic expansion.

However, MS stars above $0.5~M_\odot$ behave differently.
They have a high entropy envelope, that constitutes most of the volume of the star. Mass transfer initially eliminates this high entropy envelope, causing the star to shrink. To capture this behavior, let us denote $m_\text{c}$ as a mass of the core and $m_\text{env} = m - m_\text{c}$ as a mass of the envelope.
Writing the balance of gravity and pressure for this envelope (rather than Eq. \ref{eq:Eq68} which would apply for a uniform star), we obtain: 
\begin{equation}
    \frac{Gm_\text{c}m_\text{env}}{R^4} \propto {\rho}^{5/3} \propto {\left(\frac{m_\text{env}}{R^3}\right)}^{5/3},
    \label{eq: core and envelope}
\end{equation}
hence
\begin{equation}
    R \propto m_\text{env}^{2/3} \ .
    \label{eq: R-m dependence 2/3}
\end{equation}
This explains why these stars shrink quickly rather than expand. This radius dependence as a function of the total mass  is sharp if the envelope mass is a small fraction of the total stellar mass, but a non-negligible fraction of its volume. \citet{HjellmingWebbink1987} also present composite polytrope models for the stellar radius responding to mass loss.

To determine the dimensional coefficient in Eq.~\ref{eq: R-m dependence 2/3}, we fitted MESA curves with analytical dependencies $R(m_{\text{env}}) = R_0 ((m - m_\text{c})/(m_0 - m_\text{c}))^{2/3}$. Here $R_0$ and $m_0$ -- a radius and a mass of a star at the initial moment of a MESA simulation (on the MS). $R$ and $m$ -- a radius and a mass in the range from the moment of deviation from the MS to the end of the steep radius drop. We varied the values of the core mass until this function best matched the radius-mass relation from MESA. We found that for stars with $2~M_\odot \lesssim m_0 \leq 10~M_\odot$, where the steep decline in radius starts at $\sim 2~M_\odot$, $m_\text{c} \cong 1.3~M_\odot$ ($\sim 65~\%$). For a star with $m_0 \sim 2~M_\odot$, we obtain $m_\text{c} \cong 1.6~M_\odot$ ($\sim 80~\%$). For $m_0 \sim 1.5~M_\odot$, we find $m_\text{c} \cong 1.35~M_\odot$ ($\sim 92~\%$).

\begin{figure} [htbp]
\centering
\includegraphics[width=1.\linewidth]{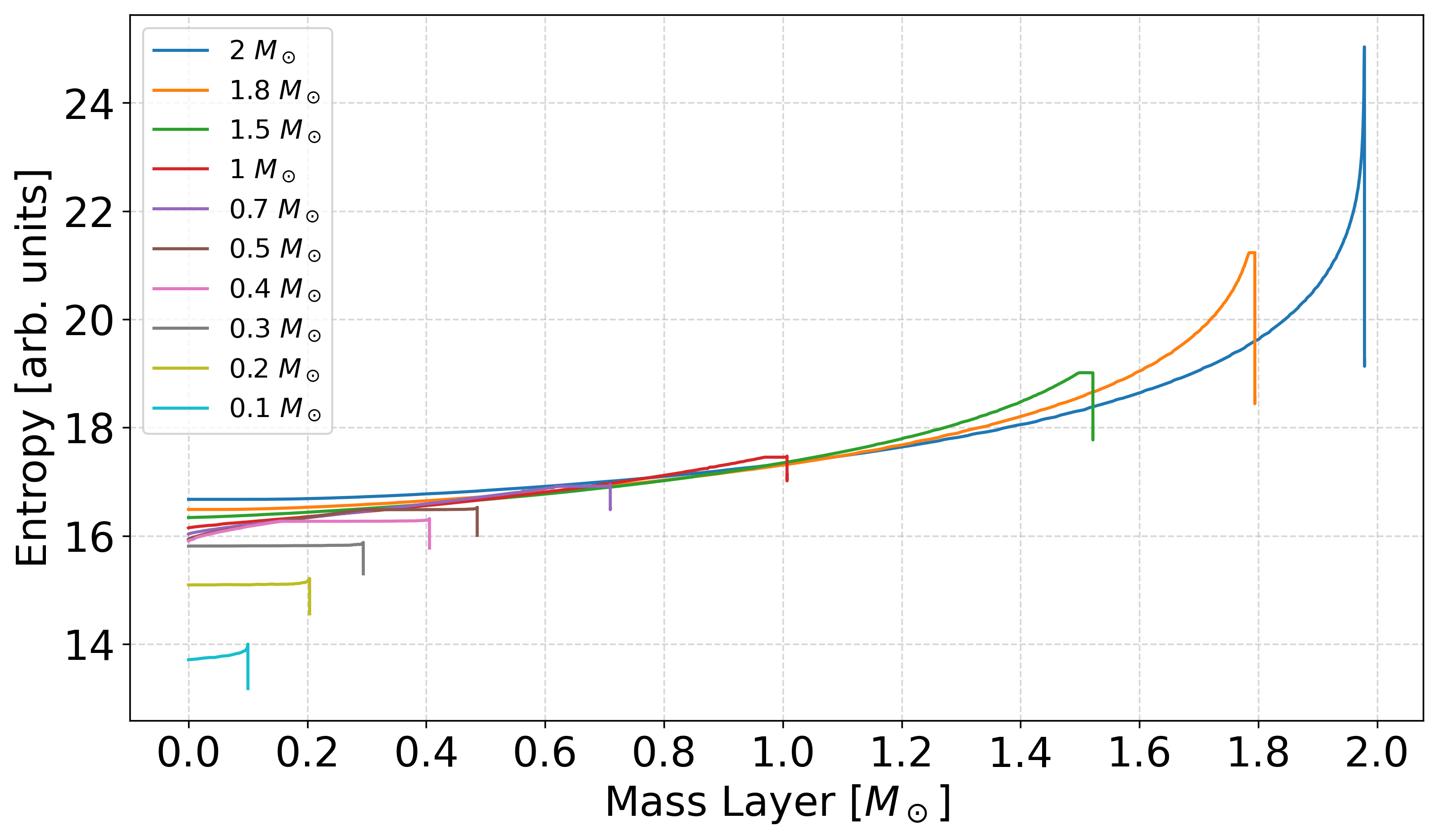}
\caption{Entropy-mass layer dependence of initially $10~M_\odot$ star when it decreases to selected masses in range from $2~M_\odot$ to $0.1~M_\odot$ ($M = 10^6~M_\odot$, $T_0 = 1$ day).}
\label{fig: Entropy-mass layer for 10 M_sun}
\end{figure}

Fig.~\ref{fig: Entropy-mass layer for 10 M_sun} demonstrates entropy-mass layer profile of the star with initial mass of $10~M_\odot$ at specific moments of time when it reaches masses in range from $2~M_\odot$ to $0.1~M_\odot$. We show that in the range $0.5~M_\odot \lesssim m \lesssim 2~M_\odot$ entropy is roughly constant over time as there is not enough time for the star to adjust itself to thermal equilibrium according to the fact that $t_\text{KH}$ is very long ($t_\text{KH} \gg t_{\dot m}$). In this region the star evolves due to GW radiation via mass loss only from the envelope. For masses below $\sim 0.5~M_\odot$ it is already not adiabatic evolution as entropy ceases to be constant and starts to decrease again. This late evolution of a star with low mass occurs at $t_\text{KH} \approx t_{\dot m}$ and the star gradually shrinks and cools down.

\begin{figure} [htbp]
\centering
\includegraphics[width=1.\linewidth]{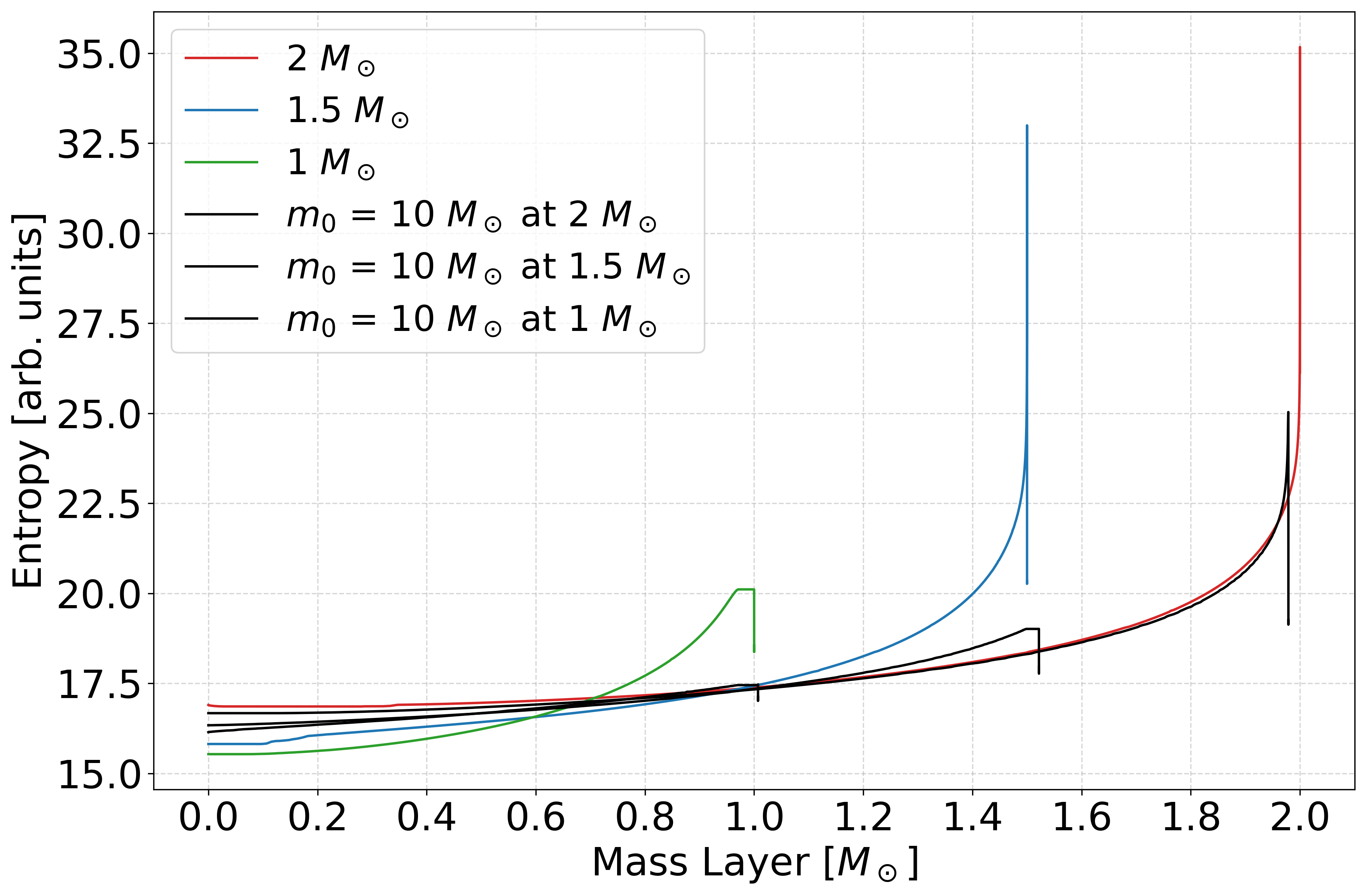}
\caption{Entropy-mass layer dependence of initially $10~M_\odot$ star when it decreases to $2, 1.5, 1~M_\odot$ and of the stars in separate simulations with these initial masses at the initial moment of time ($M = 10^6~M_\odot$, $T_0 = 1$ day).}
\label{fig: S(m) for 10 M_sun at spec. mass coordi. and 4 sep. sim. with diff. m}
\end{figure}

In Fig.~\ref{fig: S(m) for 10 M_sun at spec. mass coordi. and 4 sep. sim. with diff. m} the three entropy profiles are displayed for the initially $10~M_\odot$ star when it reaches $2, 1.5, 1~M_\odot$ and entropy profiles from three separate simulations that were run starting from these masses. When the star mass drops, for example, to $1.5~M_\odot$ in MESA simulation, its entropy profile closely follows the one from the MESA simulation when the star was at $2~M_\odot$. This is in contrast to the entropy profile from the MESA simulation of a star with initially $1.5~M_\odot$ where the entropy of its outer layers is significantly larger than in the core. 

Around $\sim 2~M_\odot$ $t_\text{KH}$ and $t_{\dot m}$ swap: $t_\text{KH}$ is very long and $t_{\dot m}$ is short and thus dominates. Therefore, the star approaches the SMBH with steeper shrinkage of the orbit due to faster mass transfer induced by the GW radiation. From the core to the outer layers it is not enough time to change entropy at almost each piece of mass and it is roughly constant. Only the entropy of a small amount of mass concentrated in the large envelope increases sharply. Once $10~M_\odot$ star reaches $\sim 0.5~M_\odot$ and evolves below it, it has enough time to adjust itself into thermal equilibrium and heat the outer layers by convection from the inner layers.

\section{Gravitational Wave Radiation}
\label{sec: Results. GW radiation}

For a binary of two point masses in circular orbit, the frequency of gravitational waves radiation in dominant quadrupole mode:

\begin{equation}
    f_\text{GW} = \frac{2}{T} = \frac{1}{\pi}\sqrt{\frac{G(m + M)}{a^3}} \ .
    \label{eq:Eq80}
\end{equation}

The characteristic strain $h_c$ is a dimensionless quantity used to assess the detectability of GW sources across the frequency spectrum. It characterizes the strength of a GW signal as a function of frequency and provides a way to compare GW sources with the sensitivity curves of detectors.

\noindent 
We used expression from \citet{Robson2019}:

\begin{equation}
    h_c(f_\text{GW}) \approx \frac{8\pi^{2/3}}{\sqrt{5}} \frac{G^{5/3}}{c^4} \frac{M_c^{5/3} f^{7/6}_\text{GW} \sqrt{T_\text{obs}}}{D} \ ,
    \label{eq:Eq82}
\end{equation}
where $D$ is distance from the detector to the binary, $T_\text{obs}$ -- observational time of the detector and $M_c$ is a chirp mass -- a magnitude describing the orbital evolution of the binary of the leading order as a result of energy loss from GW emission:

\begin{equation}
    M_c = \frac{{(mM)}^{3/5}}{{(m + M)}^{1/5}} \ .
    \label{eq:Eq83}
\end{equation}
In GW observations, the chirp mass is typically measured much more accurately than the component masses because it governs the evolution of the GW frequency. The planned LISA mission will observe GW signals, enabling measurements of the characteristic strain and inference of source parameters such as the chirp mass.

We compute the strain spectral density in order to compare our results with those of \citet{Olejak2025}. The strain spectral density \citep{Robson2019} is a frequency-dependent measure of the strength of a GW signal or a detector noise. For a gravitational-wave source, it is related to the characteristic strain by

\begin{equation}
    \sqrt{S_h(f_\text{GW})} = \frac{h_c(f_\text{GW})}{\sqrt{f_\text{GW}}} \ .
    \label{eq:Eq84}
\end{equation}

\begin{figure*} [htbp]
\centering
\includegraphics[width=\textwidth]{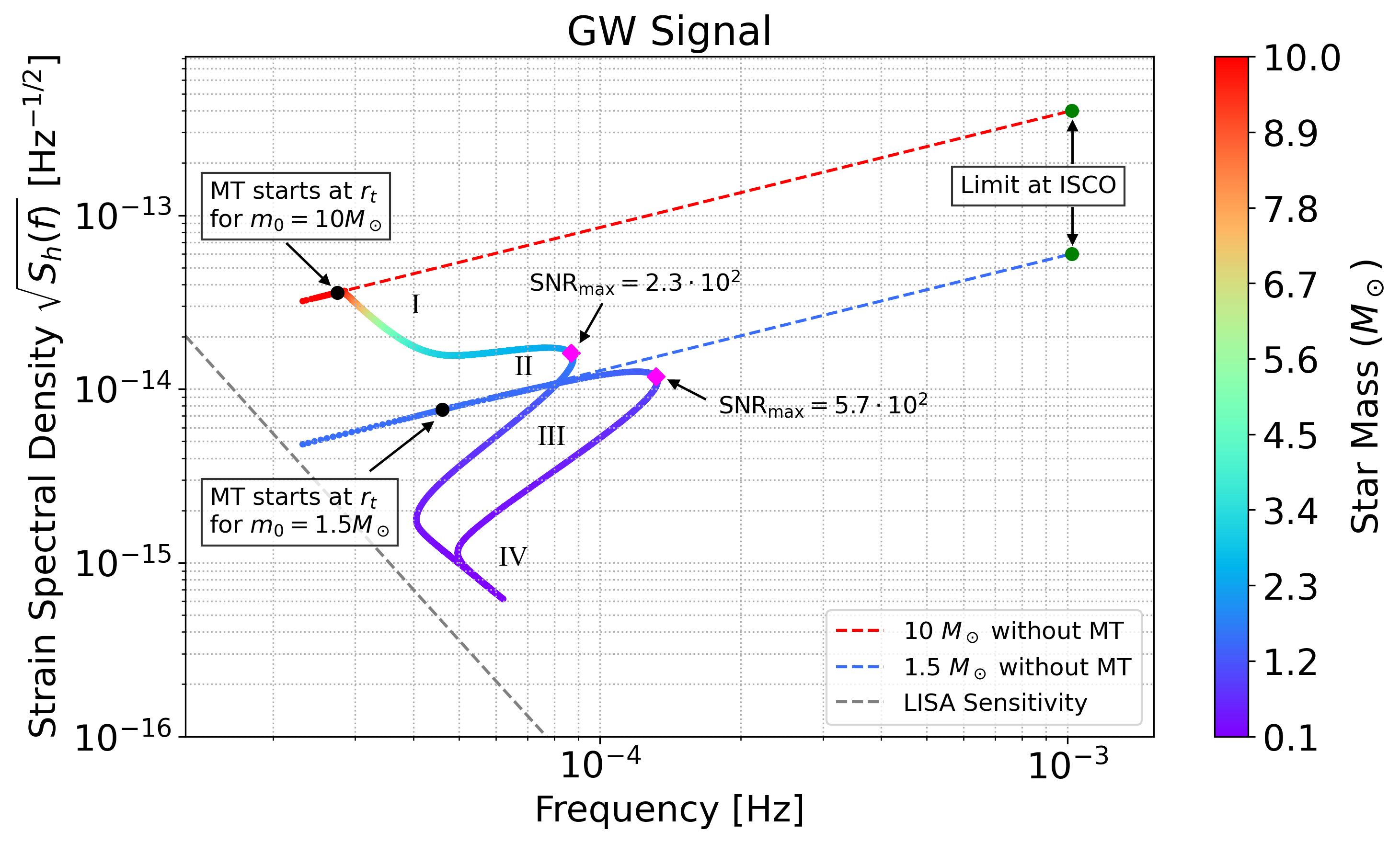}
\caption{Strain spectral density of point masses and stars approaching the $4.3 \cdot 10^6~M_\odot$ SMBH. The initial binary separation and period are $a_0 \approx 420~R_\odot$ and $T_0 = 1$ day. The distance from LISA to the binary is $D = 8$ kpc. The observational time of the LISA detector is $T_\text{obs} = 4$~years. The  LISA sensitivity curve is plotted by the gray dashed line. Point masses without mass transfer (MT) marked as a straight dashed red line for the $10~M_\odot$ and a straight dashed blue line for the $1.5~M_\odot$. Initially $10~M_\odot$ and $1.5~M_\odot$ stars transferring mass marked as the colored curves. The color bar on the right indicates the evolved star mass. Black dots mark the beginning of mass transfer ($\text{SNR}_{10 M_\odot} \approx 17$ and $\text{SNR}_{1.5 M_\odot} \approx 16$) and green dots mark limit for the two point masses inspiral at the innermost stable circular orbit (ISCO) ($\text{SNR}_{10 M_\odot} \approx 1.5 \cdot 10^6$ and $\text{SNR}_{1.5 M_\odot} \approx 2.2 \cdot 10^5$). Magenta dots mark the best detectable region corresponding to the transition between evolutionary stages II and III of initially $10~M_\odot$ star at $\sim 1.5~M_\odot$ ($\text{SNR}_\text{max} \approx 2.3 \cdot 10^2$) and initially $1.5~M_\odot$ star at $\sim 1~M_\odot$ ($\text{SNR}_\text{max} \approx 5.7 \cdot 10^2$).}
\label{fig: Strain spectral density}
\end{figure*}

In order to match the supermassive black hole residing in the Milky Way galaxy -- Sagittarius A*, we changed the mass of the SMBH in the simulation from $10^6~M_\odot$ to $4.3 \cdot 10^6~M_\odot$. We present $\sqrt{S_h(f_\text{GW})}$ for simulations of initially $10~M_\odot$ and $1.5~M_\odot$ stars on the Fig. \ref{fig: Strain spectral density}.  Black dots mark the beginning of mass transfer for each mass. For $10~M_\odot$ at this point a signal-to-noise ratio (SNR) is about $17$ and for $1.5~M_\odot$ at this point SNR is about $16$. In each case the system was evolved with the initial period of $1$ day (and $a_0 \approx 420~R_\odot$ respectively) and we took $D = 8$ kpc as the distance from the LISA to the center of the Milky Way galaxy. We used the observational time of the LISA detector $T_\text{obs} = 4$~years.

In the binary system of two point masses (a compact object and an SMBH when there is no mass transfer) -- $\sqrt{S_h(f_\text{GW})}$ will increase monotonically with the frequency growth (as the binary orbit shrinks) as shown on the graph by dashed red line for initially $10~M_\odot$ and dashed blue line for initially $1.5~M_\odot$. Green dots mark limit at $a_\text{ISCO}$ with SNR about $1.5 \cdot 10^6$ for $10~M_\odot$ and SNR about $2.2 \cdot 10^5$ for $1.5~M_\odot$.
Our results match well ones from \citet{Olejak2025} although their GW signal is stronger as they consider a more evolved star with a denser core. They also model non-conservative mass transfer, hence their system evolves faster towards higher frequencies, more compatible with the range of LISA. 

When instead of a point mass we consider a star, it overfills its Roche lobe and mass transfer starts according to the mechanisms investigated in this work. We observe deviation from straight lines downward (from red for initially $10~M_\odot$ star and from blue for initially $1.5~M_\odot$ star on Fig.~\ref{fig: Strain spectral density}) and the GW signal is emitted on lower frequencies where LISA is less sensitive. 
The signal shape corresponds to following stages of evolution. Stage I: orbit shrinks, GW frequency grows, $\sqrt{S_h(f_\text{GW})}$ decreases. At the beginning of this stage $t_{\dot m} (\propto t_\text{GW}) \approx 2.4 \cdot 10^6$~years. At the end of this stage SNR $\approx 19$ for $10~M_\odot$. Stage II: orbit shrinks drastically, GW frequency grows steeply, $\sqrt{S_h(f_\text{GW})}$ increases gradually. At the beginning of this stage $t_{\dot m} \approx 3.1 \cdot 10^6$~years. At the end of this stage SNR $\approx 2.3 \cdot 10^2$ for initially $10~M_\odot$ star at $\sim 1.5~M_\odot$ and SNR $\approx 5.7 \cdot 10^2$ for initially $1.5~M_\odot$ star at $\sim 1~M_\odot$. These dots, marked by magenta on Fig.~\ref{fig: Strain spectral density}, correspond to the best detectable region. Stage III: orbit expands, GW frequency and $\sqrt{S_h(f_\text{GW})}$ decrease significantly. At the beginning of this stage $t_{\dot m} \approx 4.7 \cdot 10^5$~years. At the end of this stage SNR $\approx 2.7$ for $10~M_\odot$ and SNR $\approx 3.1$ for $1.5~M_\odot$. Stage IV: orbit shrinks moderately, GW frequency grows moderately, $\sqrt{S_h(f_\text{GW})}$ decreases more significant than in stage I. At the beginning of this stage $t_{\dot m} \approx 1.5 \cdot 10^7$~years. At the end of this stage SNR $\approx 3.3$ for both $10~M_\odot$ and $1.5~M_\odot$.

The strain spectral density of initially $10~M_\odot$ star at stage I deviates sufficiently from the signal of two point masses coalescence as when the star starts to lose mass, its radius decreases too (evolution along the MS) as it was noted in Section \ref{sec: Results. Radius-Mass relation}. On the contrary, initially $1.5~M_\odot$ star starts already from stage II since it immediately quits the main sequence and stage I is absent for it. The strain spectral density of star with $m_0 = 1.5~M_\odot$ is almost the same as for 2 point masses binary case because in this region the star adiabatically contracts (and its radius decreases significantly) almost at constant mass, so mass loss rate is very small.

Currently, population of stars and stellar BHs with such short orbital period ($\sim 1$ day) in the Milky Way is anticipated with a probability of $\sim 10^{-2}$ \citep{Rom2024, Rom2026}.
However, if such a star does exist, \citet{Gourgoulhon2019} expect that MS stars of mass lower than $\sim 2.5~M_\odot$ orbiting Sgr A* will be detectable with SNR above 10 during one year of LISA operation, before mass transfer starts. We show that the SNR actually becomes stronger after mass transfer begins, reaching a maximum of over a hundred.
In addition, as shown by \citet{Olejak2025}, LISA would be able to detect GW signals from distances of up to $\sim 1$ Gpc if post-mass-transfer remnants (stripped stellar cores) or stellar-mass compact objects (white dwarfs, neutron stars, or stellar-mass black holes) in similarly short-period orbits around SMBHs exist in the nuclei of more distant galaxies.

\section{Summary and Discussion}
\label{sec: Summary and Discussion}

We investigated the evolution of stellar radii under mass transfer using the MESA stellar evolution code. We start with the initially $0.5-10~M_\odot$ main-sequence stars. These stars evolve in a binary system with a supermassive black hole of $10^6~M_\odot$ due to the emission of gravitational waves. The total angular momentum of the binary is conserved, so mass transfer is stable. We characterized the timescales governing mass transfer associated with gravitational-wave emission and thermal adjustment. We distinguished 4 stages of stellar evolution. We found that (I) stars with initial masses above approximately $2~M_\odot$ remain on the main sequence while gradually losing mass and contracting until their mass decreases to $\sim 2~M_\odot$. (II) Beyond this threshold, the gravitational-wave inspiral timescale becomes shorter than the thermal timescale, preventing the star from maintaining thermal equilibrium. As a result, the star leaves the MS and evolves adiabatically.
In contrast to \citet{Linial2017}, but in agreement with \citet{Dai2013b}, our results demonstrate that such stars initially contract. This behavior arises because mass loss primarily affects the star’s extended, high-entropy envelope, which contains most of the volume, while the dense core retains its entropy within a small radius.

(III) When stellar mass drops below $0.5-1~M_\odot$ (depending on the initial stellar mass), the envelope is stripped and the nearly uniform core becomes exposed. As a result, the star begins to expand adiabatically. This is also in agreement with the findings of \citet{Dai2013b}. (IV) Once the mass drops below $0.15-0.4~M_\odot$, the thermal and gravitational-wave timescales become comparable, allowing the star to cool and contract while remaining close to thermal equilibrium. This is a new regime, initially suggested by \citet{Linial2017}, that does not exist in the analysis of \citet{Dai2013} who assume adiabatic evolution.

Mass transfer slows the orbital evolution compared with the case of a compact object companion. As a result, the gravitational waves signal is weaker and evolves slower, causing such binaries to spend significant time at low frequencies close to the LISA sensitivity curve. We showed that the GW emission signal from the binary of initially $0.5-10~M_\odot$ stars and the SMBH of $4.3 \cdot 10^6~M_\odot$ is strongest in the regime where the star contracts rapidly, but in general hardly detectable by LISA.
The population of stars in the Milky Way center with such short orbital periods ($\sim 1$ day) is expected to be very low, with an estimated probability of $\sim 10^{-2}$ in the vicinity of Sgr A* at the present moment \citep{Rom2024, Rom2026}. 
In contrast, in the nuclei of more distant galaxies, post-mass-transfer remnants or stellar-mass compact objects could produce detectable signals, given that LISA is expected to be sensitive to such sources out to distances of $\sim 1$ Gpc \citep{Olejak2025}. GW signal from a compact object-SMBH binary can be detected better as the characteristic strain of GWs monotonically and significantly grows with orbital shrinkage.

\begin{acknowledgments}
This research was partially supported by GIF 3013007966 and BSF grants. AS acknowledges support from Kaete Klausner scholarship. We thank Elisha Modelevsky and Stephen Justham for helpful discussions.
\end{acknowledgments}

\bibliography{references}

\appendix
\section{Change of angular momentum and equation of motion}
\label{sec: Change of angular momentum and equation of motion}

In the conservative case in terms of angular momentum:

\begin{equation}
    \frac{\dot{L}}{L} = \frac{\dot{m}}{m} + \frac{1}{2}\frac{\dot{a}}{a} \ ,
    \label{eq:Eq85}
\end{equation}

\begin{equation}
    \frac{\dot{L}}{L}\Big|_{\text{GW}} = \frac{1}{2}\frac{\dot{a}}{a}\Big|_{\text{GW}} \ ,
    \label{eq:Eq86}
\end{equation}

\begin{equation}
    \frac{\dot{L}}{L} = \frac{\dot{L}}{L}\Big|_{\text{GW}} \rightarrow \frac{\dot{m}}{m} + \frac{1}{2}\frac{\dot{a}}{a} = \frac{1}{2}\frac{\dot{a}}{a}\Big|_{\text{GW}} \ .
    \label{eq:Eq87}
\end{equation}

At the Roche lobe (reminding that $R \propto m^{\epsilon}$):

\begin{equation}
    \frac{d}{dt}\left(\frac{R}{R_\text{RL}}\right) = 0 \rightarrow \frac{\dot{R}}{R} = \frac{\dot{R}_\text{RL}}{R_\text{RL}} \rightarrow \epsilon\frac{\dot{m}}{m} = \frac{\dot{a}}{a} + \frac{1}{3}\frac{\dot{m}}{m} \ .
    \label{eq:Eq88}
\end{equation}

Expressing $\dot{a}/a$ from Eq. \ref{eq:Eq88} and inserting it in Eq. \ref{eq:Eq87}, we end up with

\begin{equation}
    \frac{\dot{m}}{m}\left(\epsilon + \frac{5}{3}\right) = \frac{\dot{a}}{a}\Big|_{\text{GW}} \ .
    \label{eq:Eq89}
\end{equation}

From Eq. \ref{eq:Eq28} and Eq. \ref{eq:Eq30} we obtain

\begin{equation}
    \frac{\dot{a}}{a}\Big|_{\text{GW}} = -\frac{1}{4t_\text{GW}} \ .
    \label{eq:Eq90}
\end{equation}

Combining Eq. \ref{eq:Eq89} and Eq. \ref{eq:Eq90}, finally we get the relation Eq. \ref{eq:Eq72}

\begin{equation}
    \frac{m/\dot{m}}{t_\text{GW}} = -4\left(\epsilon + \frac{5}{3}\right) \ .
    \label{eq:Eq91}
\end{equation}

\end{document}